\documentclass{webofc}
\usepackage[varg]{txfonts}   % Web of Conferences font
\usepackage{hyperref}
\usepackage{url}
\usepackage{multicol}
\hypersetup{colorlinks=true,citecolor=blue,urlcolor=blue,linkcolor=blue}
\makeatletter
\renewcommand\section{%
  \@startsection {section}{1}{\z@}%
  {-9pt plus-2pt minus-2pt}%
  {4pt plus2pt minus2pt}%
  {\@SectionFont\raggedright}%
}
\renewcommand\subsection{%
  \@startsection{subsection}{2}{\z@}%
  {-9pt plus-2pt minus-2pt}%
  {4pt plus2pt minus2pt}%
  {\@SubsectionFont\raggedright}%
}
\makeatother

\begin{document}
\title{Modeling Gamma-Ray Bursts Using CRISP}

\author{\firstname{Leonel} \lastname{Morejon}\inst{1}\fnsep\thanks{\email{leonel.morejon@uni-wuppertal.de}} \and
        \firstname{Therese} \lastname{Paulsen}\inst{1}\fnsep\thanks{\email{paulsen@uni-wuppertal.de}} \and
        \firstname{Karl-Heinz} \lastname{Kampert}\inst{1}\fnsep\thanks{\email{kampert@uni-wuppertal.de}}
        % etc.
}

\institute{Bergische Universität Wuppertal, Gaußstraße 20, 42119 Wuppertal
          }

\abstract{Astrophysical neutrinos are a clear identifier of photohadronic interactions in sources of Ultra-High-Energy Cosmic Rays (UHECRs). As the Pierre Auger Observatory has shown that UHECRs exhibit a composition heavier than protons at Earth, we expect that these heavy primaries undergo nuclear cascades due to photodisintegration in the dense source environments from which they originate. In this contribution, we will derive the neutrino spectra including those originating from nuclei undergoing photohadronic interactions. This will be explored in the context of modeling the emission region of a gamma-ray burst. Photohadronic interactions will be modeled using the new framework for Cosmic Ray Stochastic Interactions for Propagation (CRISP), which employs an analytic approach to compute the underlying probabilistic description of UHECR interactions. 
}
\maketitle
\section{Introduction}
\label{intro}

Gamma-ray bursts (GRBs) have long been considered a natural source of UHECRs, because of their high emission of $\gamma$-rays, with some of the first predictions dating back to 1995 \cite{1995PhRvL..75..386W,1995ApJ...453..883V}. Recently, the IceCube collaboration demonstrated a lack of coincidences between GRBs and very high-energy neutrinos \cite{Abbasi:2022whi}, placing stringent constraints on the neutrino fluences and GRB modeling parameters. However, observing neutrinos in the high-energy range is important not only for identifying potential UHECR sources, but also for pinpointing GRB emission mechanisms \cite{Pitik_2021}. In this work we examine a GRB jet, described by the Internal Shock (IS) model, with an iron primary undergoing nuclear cascades from photodisintegration in the dense source environment. Simulating this time-dependent composition evolution has been achieved for full nuclear cascades using the NeuCosmA software \cite{2018A&A...611A.101B,2018ApJ...854...54R}. The recently introduced framework CRISP \cite{2026A&A...708A..21M} distinguishes itself from NeuCosmA by describing nuclear cascades analytically, as Markov jump processes, rather than through time integration of a system of ordinary differential equations for the density of each species. A detailed discussion on the equivalence between these approaches can be found in \cite{2026A&A...708A..21M}.

%\cite{2026A&A...708A..21M} discusses the equivalence between these approaches in detail.
\section{GRB modeling with CRISP}
\label{sec-1}
\begin{figure}[h!]
\centering
\sidecaption
\includegraphics[width=8cm,clip]{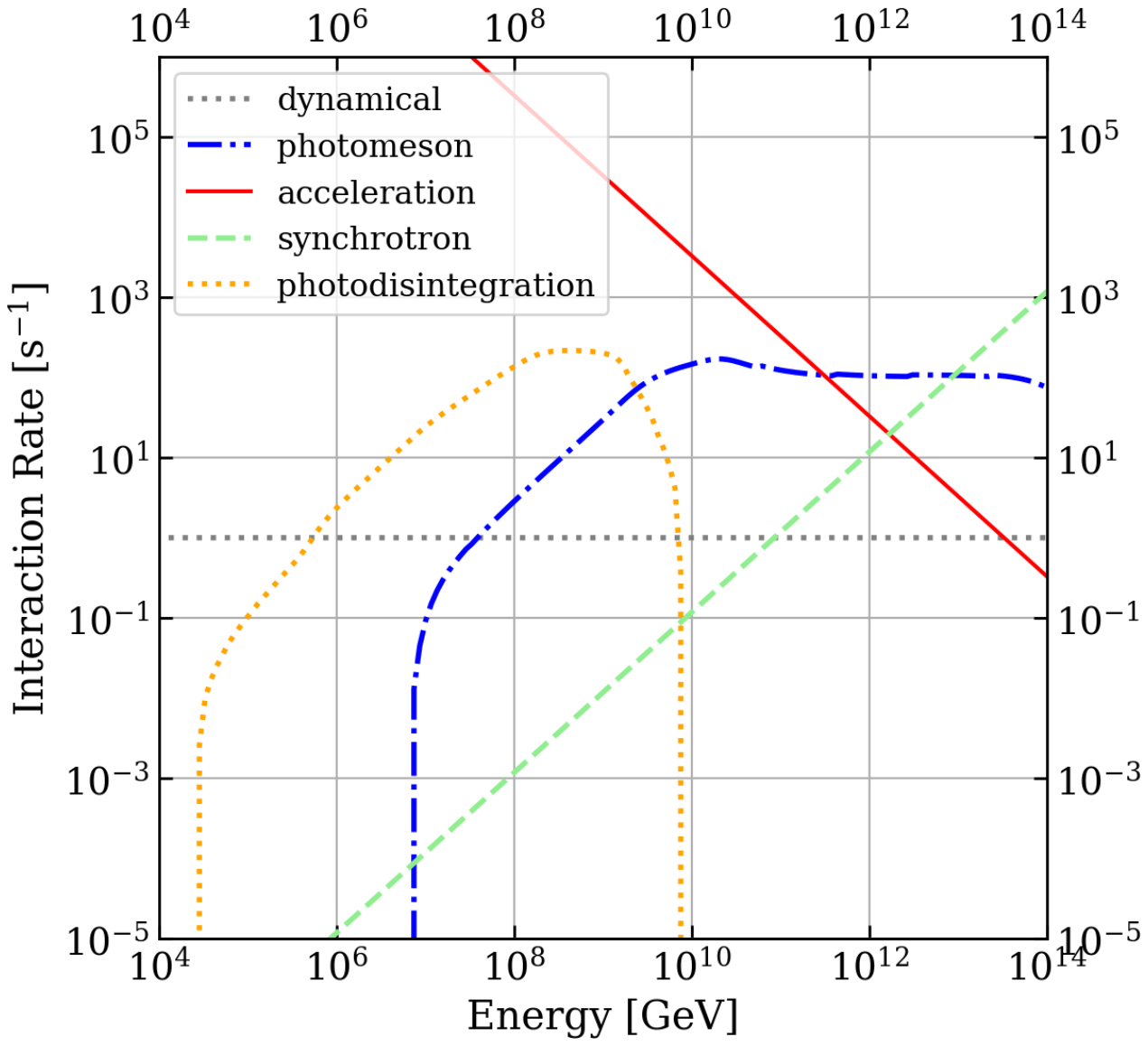}
\caption{Interaction rates for a primary iron nucleus, $^{56}$Fe, in the emission region of the GRB jet following the IS model: photodisintegration and photomeson production (stochastic losses), synchrotron and adiabatic/dynamical losses (continuous losses), and the acceleration rate. Stochastic losses dominate below $\sim10^{13}\mathrm{GeV}$, where synchrotron takes over. The acceleration rate crosses the total loss rate, set by photomeson interactions at that energy, near $E_{\rm max}\sim3\times10^{11}~\mathrm{GeV}$.}
\label{fig:interaction_rates}
\end{figure}

The GRB scenario considered here is the optically thick case of \cite{2018A&A...611A.101B}, \textit{i.e.} the source is optically thick to $A\gamma$ interactions. We employ the modeling parameters of Table~2 in \cite{2019JCAP...11..007M}: a jet with Lorentz factor $\Gamma = 300$, luminosity $L_{\gamma} = 10^{53}~\mathrm{erg~s^{-1}}$, duration $T_{90} = 10~\mathrm{s}$, collision radius $R_{c} = 2 \times 10^{8}~\mathrm{km}$, baryonic loading $\xi = 10$, and a seed photon spectrum following a broken power law with spectral indices $p_{1} = -1$ and $p_{2} = -2$ before and after the break, respectively. The low-energy cutoff at $\epsilon_{\gamma,\mathrm{min}}^{\prime} = 100~\mathrm{eV}$ suppresses photodisintegration at the highest energies, where photomeson production becomes dominant (Fig.\ref{fig:interaction_rates}). Equating the total energy-loss rate to the acceleration rate gives a maximum energy for this iron of $E_{\mathrm{max}} \sim 3\times 10^{11}\mathrm{GeV}$. We employ a one-zone geometry in which the collision radius is $R_{c} = 2\Gamma^{2} t_{v}/(1+z)$ and the shell width is $\Delta d^{\prime} = \Gamma c t_{v}/(1+z)$; the photon spectrum normalization is $u_{\gamma}^{\prime} = L_{\gamma}/(4\pi \Gamma^{2}R^{2})$, and $B^{\prime}$ follows from equipartition. 

\section{Source Emission Spectra}
\label{sec-2}
\begin{figure*}[h!]
\centering
\includegraphics[width=6.3cm,clip]{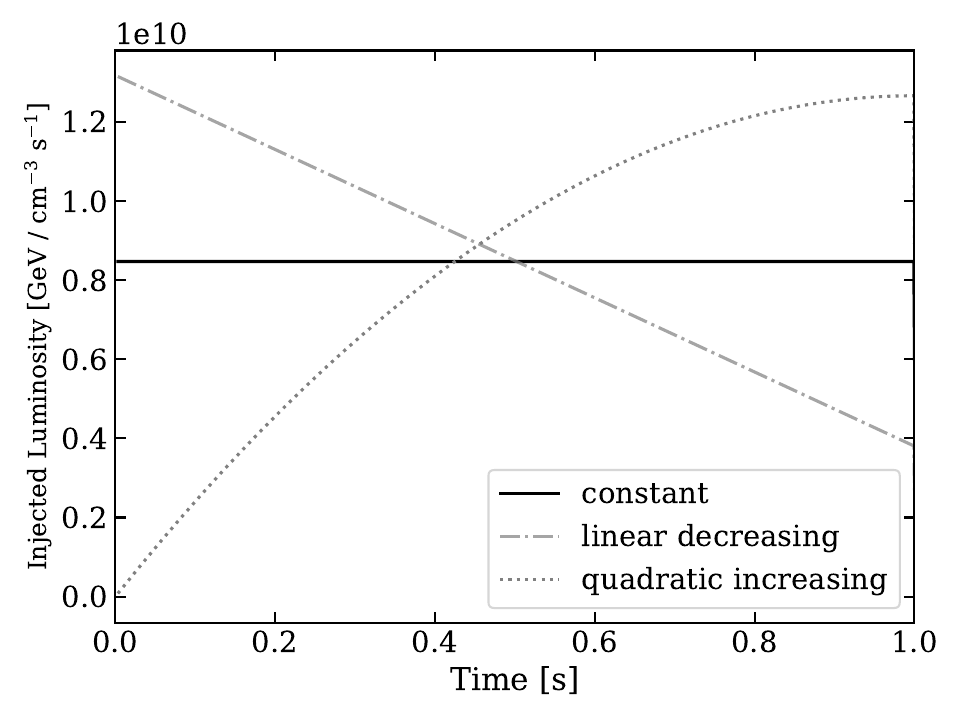}
\includegraphics[width=6.5cm,clip]{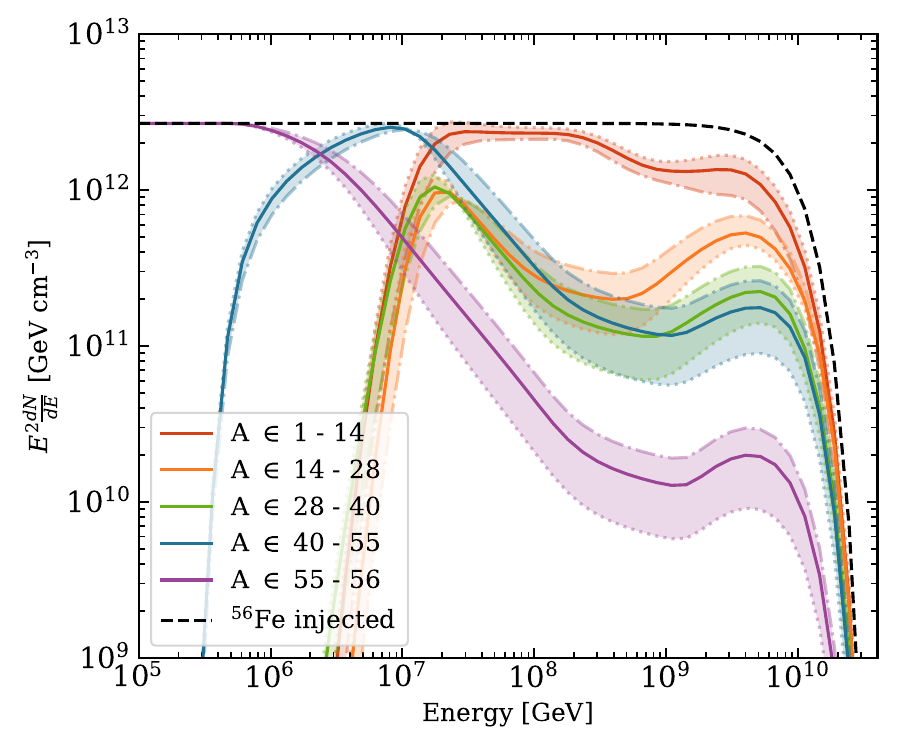}
\caption{Left: Temporal dependence of the injected luminosity: constant (solid line), linearly decreasing (dash-dotted), and quadratically increasing (dotted).
Right: Nuclear SEDs for different mass groups, including the injected primary $^{56}$Fe (black dashed). The disintegration products are grouped into mass ranges (see legend). The solid, dotted and dashed lines correspond to the different injection scenarios shown in the left graph. Taken from \cite{2026A&A...708A..21M}.}
\label{fig2}
\end{figure*}

Following \cite{2026A&A...708A..21M}, we calculate the time-dependent spectral densities for each nuclear species within the source. The final yield for each UHECR species, $\textbf{N}(\gamma, t_{\mathrm{inj}})$, is
\begin{equation}
 \textbf{N}(\gamma, t_{\mathrm{inj}}) = \int^{t_{\mathrm{inj}}}_{0} \bf{\tilde{Q}}^{\mathrm{ext}}  (\gamma,t^{\prime}) {\textbf{P}}^{\,t_{\mathrm{inj}}-t^{\prime}} (\gamma) \mathrm{d}t^{\prime} \: ,
 \label{eq:inj_vec}
\end{equation}
\noindent where $\bf{\tilde{Q}}^{\mathrm{ext}}$ is the injection rate vector, ${\textbf{P}}$ is the transition probability between nuclear species, $\gamma$ is the Lorentz boost, and $t_{\mathrm{inj}}$ is the injection time. ${\textbf{P}}$ is time-independent when continuous energy losses are subdominant, as in the case studied here (Fig.\ref{fig:interaction_rates}); \cite{2026A&A...708A..21M} details the general procedure otherwise. Here, Eq.\eqref{eq:inj_vec} is solved for $^{56}$Fe injection with a power-law dependence on the boost and an exponential cutoff. Figure~\ref{fig2} (left) shows three injection time profiles -- constant, linearly decreasing, and quadratically increasing -- parametrized to give the same total injected luminosity, and (right) the resulting nuclear SEDs for each case. Nuclei are assumed to escape after a characteristic time/distance (advective escape). The resulting nuclear densities agree with the independent calculations of \cite{2019JCAP...11..007M} and \cite{2018A&A...611A.101B} to within $15$-$30~\%$ on the proton, neutron, and intermediate-mass peaks.

\begin{figure}[h!]
\centering
\includegraphics[width=12cm,clip]{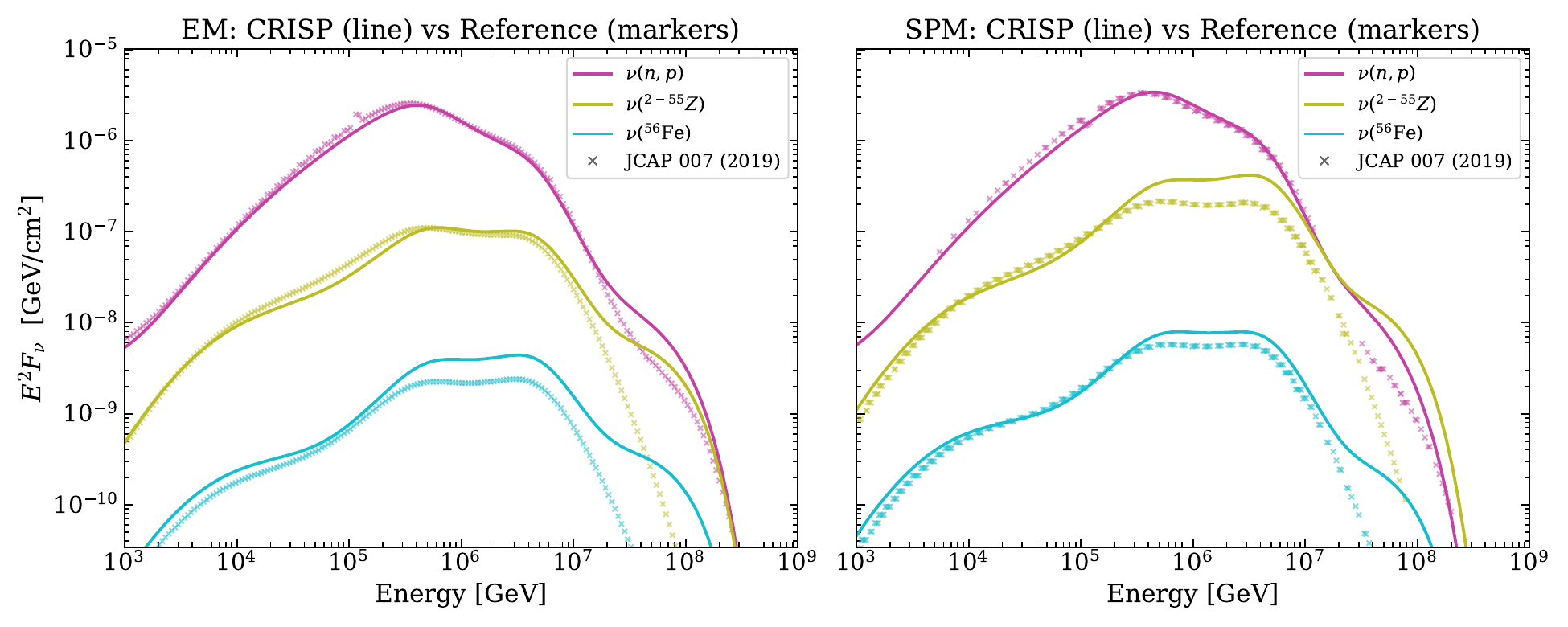}
\caption{Neutrino fluence per shell from the GRB source of Sect.~\ref{sec-1}, computed with CRISP (solid lines) for the Empirical Model (left) and the Single Particle Model (right), compared against the values traced directly from Fig.~9 (lower left) of \cite{2019JCAP...11..007M} (markers, labeled JCAP 007 (2019)). Colors indicate the parent species whose photomeson interaction produced the pion: secondary nucleons $\nu(n,p)$, intermediate nuclei $\nu(^{2-55}Z)$, and the surviving primary $\nu(^{56}\mathrm{Fe})$.}
\label{fig:nu1}
\end{figure}

\medskip
\noindent
The accompanying neutrinos produced by the photohadronic interactions in this source are obtained by following the photomeson kernel and decay-chain relations of \cite{2010ApJ...721..630H}: charge- and species-resolved kernels give the pion and kaon yields, propagated through the full $\pi/K \rightarrow \mu \rightarrow e$ decay chains (including muon helicity) to obtain the neutrino spectra. Figure~\ref{fig:nu1} compares the resulting spectra, computed with the Empirical Model (EM, left) and the Single Particle Model (SPM, right), against the values traced directly from \cite{2019JCAP...11..007M}. As in \cite{2019JCAP...11..007M}, the EM output falls below the SPM in every channel, a consequence of its $A^{2/3}$ nuclear pion-yield suppression. Agreement between this work and \cite{2019JCAP...11..007M} is excellent for the dominant nucleon channel, $\nu(n,p)$, for both models (within $2~\%$ for the SPM and $3~\%$ for the EM). The subdominant intermediate-mass, $\nu(^{2-55}Z)$, and $^{56}$Fe channels agree less tightly with the corresponding model in \cite{2019JCAP...11..007M}, within a factor of $\sim2$ for both the SPM and the EM, with no systematic excess or deficit in either. This wider spread reflects differences between the two independent implementations, such as the nuclear cross-section data and the specific set of nuclear species tracked by each, rather than a shortcoming of either model; since these channels sit one to three orders of magnitude below the nucleon peak, it has little effect on the total neutrino output. The small excess visible above $\sim3\times10^{7}$~GeV in all channels is due to the addition of neutrinos from kaon-decay, which are absent from the photomeson model of \cite{2019JCAP...11..007M}; excluding kaons, all channels agree with \cite{2019JCAP...11..007M} to within order unity across the full energy range.

%it has little effect on the total neutrino output. The small excess visible above $\sim3\times10^{7}$~GeV in all channels is due to the kaon-decay neutrinos included here, which are absent from the photomeson model of \cite{2019JCAP...11..007M}; excluding kaons, all channels agree with \cite{2019JCAP...11..007M} to within order unity across the full energy range.

\section{Conclusions}
\label{sec-3}
We have applied the CRISP framework to model UHECR and neutrino emission from a GRB internal-shock source in the optically thick regime, following the parameters of \cite{2018A&A...611A.101B,2019JCAP...11..007M}. Using the interaction rates for a primary $^{56}$Fe nucleus, we determined the source's maximum attainable energy and showed that stochastic energy losses dominate over continuous losses up to $\sim10^{13}~\mathrm{GeV}$, justifying the time-independent transition probability used in Eq.\eqref{eq:inj_vec}. The baryon number is conserved to better than $0.1\%$ throughout the calculation. We also computed nuclear SEDs under different temporal injection profiles, showing that CRISP's probabilistic approach directly accommodates time-dependent injection.

\medskip
We further computed the neutrino spectra using both the SPM and EM photomeson models. The resulting spectra agree with the independent, NeuCosmA-based calculation of \cite{2019JCAP...11..007M} to within $\lesssim3\%$ in the peak nucleon-channel fluence for both models (Fig.~\ref{fig:nu1}), validating CRISP's neutrino production pipeline against an established reference in the literature. Future work will extend this approach to other source classes and combine the in-source treatment presented here with extragalactic propagation within a single, consistent framework.

%\medskip
\section*{Acknowledgements}
This work was supported by BMBF Verbundforschung Astroteilchenphysik (Project No. 05A23PX1).
%\twocolumn
\begin{multicols}{2}
\bibliography{references.bib} 

@article{Abbasi:2022whi,
    author = "Abbasi, R. and others",
    title = "{Searches for Neutrinos from Gamma-Ray Bursts Using the IceCube Neutrino Observatory}",
    doi = "10.3847/1538-4357/ac9785",
    journal = "Astrophys. J.",
    volume = "939",
    number = "2",
    pages = "116",
    year = "2022"
}

@ARTICLE{1995ApJ...453..883V,
       author = {{Vietri}, Mario},
        title = "{The Acceleration of Ultra-High-Energy Cosmic Rays in Gamma-Ray Bursts}",
      journal = {apj},
         year = 1995,
        month = nov,
       volume = {453},
        pages = {883},
          doi = {10.1086/176448},
       adsurl = {https://ui.adsabs.harvard.edu/abs/1995ApJ...453..883V}
}

@ARTICLE{2010ApJ...721..630H,
       author = {{H{\"u}mmer}, S. and {R{\"u}ger}, M. and {Spanier}, F. and {Winter}, W.},
        title = "{Simplified Models for Photohadronic Interactions in Cosmic Accelerators}",
      journal = {apj},
         year = 2010,
       volume = {721},
       number = {1},
        pages = {630},
          doi = {10.1088/0004-637X/721/1/630},
       adsurl = {https://ui.adsabs.harvard.edu/abs/2010ApJ...721..630H}
}

@ARTICLE{1995PhRvL..75..386W,
       author = {{Waxman}, Eli},
        title = "{Cosmological Gamma-Ray Bursts and the Highest Energy Cosmic Rays}",
      journal = {prl},
         year = 1995,
        month = jul,
       volume = {75},
       number = {3},
        pages = {386-389},
          doi = {10.1103/PhysRevLett.75.386},
       adsurl = {https://ui.adsabs.harvard.edu/abs/1995PhRvL..75..386W}
}

@ARTICLE{2018A&A...611A.101B,
       author = {{Biehl}, D. and {Boncioli}, D. and {Fedynitch}, A. and {Winter}, W.},
        title = "{Cosmic ray and neutrino emission from gamma-ray bursts with a nuclear cascade}",
      journal = {aap},
         year = 2018,
        month = apr,
       volume = {611},
          eid = {A101},
        pages = {A101},
          doi = {10.1051/0004-6361/201731337},
       adsurl = {https://ui.adsabs.harvard.edu/abs/2018A&A...611A.101B}
}

@article{Pitik_2021,
doi = {10.1088/1475-7516/2021/05/034},
url = {https://dx.doi.org/10.1088/1475-7516/2021/05/034},
year = {2021},
month = {may},
publisher = {IOP Publishing},
volume = {2021},
number = {05},
pages = {034},
author = {Pitik, Tetyana and Tamborra, Irene and Petropoulou, Maria},
title = {Neutrino signal dependence on   gamma-ray burst emission mechanism},
journal = {Journal of Cosmology and Astroparticle Physics}
}

@ARTICLE{2026A&A...708A..21M,
       author = {{Morejon}, Leonel and {Kampert}, Karl-Heinz},
        title = "{Stochastic analysis of ultrahigh-energy cosmic ray interactions}",
      journal = {aap},
         year = 2026,
        month = mar,
       volume = {708},
          eid = {A21},
        pages = {A21},
          doi = {10.1051/0004-6361/202557405},
       adsurl = {https://ui.adsabs.harvard.edu/abs/2026A&A...708A..21M}
}

@ARTICLE{2019JCAP...11..007M,
       author = {{Morejon}, L. and {Fedynitch}, A. and {Boncioli}, D. and {Biehl}, D. and {Winter}, W.},
        title = "{Improved photomeson model for interactions of cosmic ray nuclei}",
      journal = {jcap},
         year = 2019,
        month = nov,
       volume = {2019},
       number = {11},
          eid = {007},
        pages = {007},
          doi = {10.1088/1475-7516/2019/11/007},
       adsurl = {https://ui.adsabs.harvard.edu/abs/2019JCAP...11..007M}
}

@ARTICLE{2018ApJ...854...54R,
       author = {{Rodrigues}, Xavier and {Fedynitch}, Anatoli and {Gao}, Shan and {Boncioli}, Denise and {Winter}, Walter},
        title = "{Neutrinos and Ultra-high-energy Cosmic-ray Nuclei from Blazars}",
      journal = {apj},
         year = 2018,
        month = feb,
       volume = {854},
       number = {1},
          eid = {54},
        pages = {54},
          doi = {10.3847/1538-4357/aaa7ee},
       adsurl = {https://ui.adsabs.harvard.edu/abs/2018ApJ...854...54R}
}
\end{multicols}
\end{document}